\documentstyle[12pt,aasms]{article}
\tighten
\begin{document}
\newcommand{\be}{\begin{equation}}
\newcommand{\ee}{\end{equation}}
\newcommand{\bea}{\begin{eqnarray}}
\newcommand{\eea}{\end{eqnarray}}
\title{The Effects of Inhomogeneities on  Evaluating \\
the Deceleration Parameter q$_0$}
\author{ R. Kantowski,  T. Vaughan, and D. Branch \footnote[1]
{Kantowski@phyast.nhn.uoknor.edu}}
\affil{
University of Oklahoma, Department of Physics and Astronomy,\\
Norman, OK 73019, USA
}

\begin{abstract}
\vskip .2 truein
Analytic expressions for distance-redshift relations which have been
corrected for the effects of  some inhomogeneities in the
Friedmann-Robertson-Walker (FRW) mass
density are used to illustrate the significance of inhomogeneities
on a determination of $q_0$ made by using Type Ia supernovae.  The value
of $q_0$ inferred from a given set of observations depends on the fractional
amount of matter in inhomogeneities and is up to 50 percent larger than that
obtained by using the standard Mattig $m$-$z$ result for
pure dust FRW models.
\end{abstract}


\section{Introduction}
\label{sec:intro}
   Some years ago one of us (Kantowski 1969, henceforth referred to as K1)
used the ``Swiss-cheese"
models to study the effects of local inhomogeneities in the FRW mass density
on the propagation of light through an otherwise homogeneous and isotropic
universe.
The Swiss-cheese models are standard pressure-free FRW models
(sometimes called dust models) from which
non-overlapping comoving spheres of zero-temperature  material
(called dust)  have been removed
and each replaced by a mass  concentrated at its center. At a given
epoch  the model looks like  Swiss-cheese full of perfectly round
holes each with a seed (representing a spherical mass) at its center
(see Fig. 1). If the
holes are small compared to the radius of the FRW universe, the mass
of the seed differs indistinguishably from the integrated mass of the
dust missing from
the sphere (see \cite{KR}). The comoving
spheres expand/contract right along with the universe itself. The
gravity field
inside each expanding sphere is given by a Schwarzschild metric of
concentrated mass. The importance of Swiss-cheese  models is that
they are explicit exact solutions to GR
(with all boundary conditions being satisfied). In such  models, effects
of local inhomogeneities on observable relations can be calculated free
of superposition approximations.

The large scale structure of the
universe, i.e., which pressure free FRW model we are actually living
in, is clearly
unaffected by how the clumps are distributed in number density or
size. The FRW structure is fixed by giving the Hubble ($H_0$) and the
deceleration  ($q_0$) parameters.  However, as we will show,
observational determination
of $q_0$ is critically affected by the clumps. Most
observations
involve collecting radiation from a finite-sized distant
source with  a much smaller ($\sim$ point) receiver. The cross
section of the  radiation (the beam) ultimately received is
consequently of the same order as the source's size. It is this size
that is important in deciding what matter is to be ascribed to the
seeds
and what to the cheese. Matter whose size is small
compared
to the  beam (e.g. an atom)  is included in the cheese (the homogeneous
dust) and
matter whose size
is large compared to the beam (e.g. a galaxy) is put into the seeds (the
inhomogeneous clumps). The reader should remember that this  is a model
which is intended to mimic the optical effects of matter and not
its actual physical distribution. For example atomic gas clouds are not
physically confined away from galaxies as is the cheese; however, the
optical effects of the gas should be independent of its location (see
Section \ref{sec:optics}).
In this model matter the same size as
the beam clearly causes trouble  and has to be dealt
with separately for each
observation. The above separation of mass based on size is not
completely well
defined because of the changing beam size from source to receiver.
Star clusters
for example would be considered  small near a galactic sized source
but large near us.

In Section \ref{sec:optics}, optical propagation equations,
originally derived in \cite{KR}, are restated. These equations
describe the propagation of a beam of electromagnetic radiation
traversing a clumpy universe. The reader is directed to \cite{KR}
 for details and to Sections 4.5.2 and 4.5.3 of
\cite{SEF} for a complete review of more recent work.
In the Appendix
it is  concluded that the net effect of the seeds is negligible for
$z < 1$ (except for uncommon lensing type situations) and in Section
\ref{sec:b=0} it is shown that the resulting $b = 0$
propagation  equation for the cross sectional area of the beam is the
associated Legendre equation.\footnote[2]{Before this manuscript could be
completed and submitted and unbeknownst to us, equation (\ref{aLe}) and its
solutions (\ref{answerL}) and (\ref{answerH}) were published by  \cite{SS}.
They deserve credit for first recognizing (\ref{Ab0}) as the Associated
Legendre equation and giving its general solution. They also show that the
general solution reduces to previously known special solutions for appropriate
parameter values.  We give our version of the solution for completeness only.
The conjectured extension of the validity of the $b=0$ equation beyond the
Swiss-Cheese models is argued by \cite{SEF} in Sections 4.5.2 and 4.5.3.}

The main result of this paper is the clear presentation of the corrected
luminosity distance [expression (\ref{answerL}) combined with
(\ref{Etherington})] and the clear demonstration of the necessity of its use
when evaluating $q_0$.
In Section \ref{sec:conclusions}  the importance of taking into
account the
effects of inhomogeneities is demonstrated by showing
how
a recent determination of $q_0$ (made by applying the magnitude-redshift
relation to SN1992bi) will be
altered.

\section{Swiss-cheese Optics}
\label{sec:optics}
Observable relations such as magnitude-redshift or angular
size-redshift are determined from expressions for the cross sectional
area $A$ of a light beam that
propagates between source and  observer.
In \cite{KR} an integral-differential
equation for the average  area $A$  as it propagates
through a Swiss-cheese universe (see Fig. 1) was given as
\be
{\sqrt{A}^{''}\over \sqrt{A}} + {\langle\xi^2\rangle\over A^2} =
{\rho_D\over \rho_F}{\cal R}_F\ ,
\label{Af}
\ee
where
\bea
{'}\ &\equiv& {d\over dv} = -{H_0\over c}(1+z)^3(1+2 q_0 z)^{1/2}{d \
\over d z}\ ,
\label{affine}\\
 \langle\xi^2\rangle &=& {\rho_I\over \rho_F} b^3\int_0^v A^2(1+ z)^6
d v\,,
\label{xi}\\
{\cal R}_F &=& -{4\pi G\over c^2}\rho_F(0)(1+z)^5\nonumber\\
&=&-3 q_0{H_0^2\over c^2}(1+z)^5\ ,
\label{RF}
\eea
and where Weinberg sign conventions have been used  (\cite{MTW}).
In (\ref{Af}), $\rho_F$ is the usual FRW mass density (the local mass
density of the cheese) and in (\ref{RF})  $\rho_F(0)$ is its current
value. The other two mass densities,  $\rho_D$ and $\rho_I$, are the
components of  $\rho_F$ made up of dust and inhomogeneities
respectively,  $\rho_F = \rho_D + \rho_I$. In the Swiss-cheese model
$\rho_I$ is given by summing all clumped masses in a large volume
(containing a representative number of clumps) and dividing by that
volume. The density  $\rho_D$ is similarly computed by summing only
the dust mass. The two driving terms for the area $A$ in (\ref{Af})
are  proportional to ${\cal R}_F$ and $\langle\xi^2\rangle$ which
represent average converging effects on the
beam as it passes respectively \underbar{through} dust  and
\underbar{near} clumps. To arrive at  (\ref{Af}) a large number of
`small' encounters with clumps had to be assumed. Any single
encounter that causes a `large' fractional change in the beam's
observed cross sectional area was excluded. Such encounters cause
lensing effects and must
necessarily be accounted for by other means  (\cite{BR}).
The $v$ parameter is  a \underbar{particular} affine parameter
which has been expressed in (\ref{affine}) as a differential operator
in redshift  $z$. Equations (\ref{affine}), (\ref{xi}), and
(\ref{RF}) are slightly simpler than the corresponding equations in
\cite{KR} because of the choice of the particular affine parameter
made here (note that no loss of physical generality is made by such a
choice).
The electromagnetic wave starts from a point on the source at $z = z_s$
where
$v =0$, and expands at first as a small part of a spherical wave
front, e.g. as a disk. As the spherical wave front encounters a clump
the disk is distorted into an ellipse. The   $ \langle\xi^2\rangle$
term in (\ref{Af}) is the effect on the wave front's area caused by
repeated distortions of the clumps.
The parameter $b$ measures the strength of the distortions and has
units of (length)$^{-1}$. Its value is defined by
\be
b^3 = 12 q_0 {H_0^2\over c^2}{\langle(G m/c^2)^{4/3}\rangle \over
\langle(G m/c^2)\rangle }\left[{(G m/c^2)^{1/3}\over a}\right]^2,
\label{b3}
\ee
which makes $b$ the order of the cube root of
the Schwarzschild radius of a typical clump divided by its opaque
radius squared (i.e., by $a^2$) and by the Hubble radius of the
universe squared (i.e., by $c^2/H_0^2$).

The reader is directed to \cite{KR} for details and assumptions made
in all derivations, including the somewhat unphysical assumption that
$ m/a^3$ is the same for all clumps. If $b = 0$, (\ref{Af}) is easily
seen to be a linear homogeneous second-order equation in  $\sqrt{A}$
[ i.e., see (\ref{Ab0})]; however, if $b\ne 0$, upon differentiation,
 (\ref{Af}) becomes a  linear homogeneous third-order equation in
$A$,

\be
A''' - 4\left[{\rho_D\over\rho_F}{\cal R}_F \right]A' - 2\left[
{\rho_D\over\rho_F}{\cal R'}_F - {\rho_I\over\rho_F}
b^3(1+z)^6\right] A = 0.
\label{A3}
\ee
The area $A$ in (\ref{Af}) and (\ref{A3}), when evaluated at the
observer, is related to the luminosity distance $D_{\ell}$ by $\delta
\Omega D_{\ell}^2=A|_0(1+z_s)^2$ or equivalently $\delta\theta
D_{\ell}=\sqrt{A}|_0 (1+z_s)$.  The luminosity distance is defined in
terms of the flux received ${\cal F}$ and the power ${\delta\cal P}$
radiated into solid angle $\delta\Omega$ by ${\cal F} = {\delta\cal
P}/\delta\Omega D_{\ell}^2$.
The solid and plane angles are constants measured at the source
($z=z_s \leftrightarrow v=v_s=0$) for a wave front expanding into
the future. The initial conditions for (\ref{Af}) are $\sqrt{A}|_s=0$
and $d \sqrt{A}/dv|_s= \delta\theta(1+z_s)$ (or equivalently $d
\sqrt{A}/dz|_s= -\delta\theta c/[H_0 (1+z_s)^2\sqrt{1+2q_0z_s}])$.
The additional boundary condition needed for (\ref{A3}) comes from
(\ref{xi}) and is $d^2\sqrt{A}/dv^2|_s= 0$. These boundary conditions
are a
consequence of our choice of affine parameter $v$ in (\ref{affine}),
so chosen to simplify  ``apparent size" boundary conditions (see the
next section).
The general power series solution
with $ b\ne 0$ is:
\bea
\sqrt{A}|_0/\delta\theta &=& D_{\ell}/ (1+z)={c\over H_0}
\Biggl\{
z -{1\over2}\Biggl[1+q_0\Biggr] z^2
+{1\over2}\left[1+{\rho_I\over\rho_F}q_0+q_0^2\right] z^3\nonumber\\
&-&{1\over2}\left[
1+{\rho_I\over\rho_F}q_0-\left({1\over4}-{3\over2}{\rho_I\over\rho_F}
\right)q_0^2+{5\over4} q_0^3 +{1\over 30}b^3{c^3\over
H_0^3}{\rho_I\over\rho_F}
\right] z^4
+O[z^5]
\Biggr\}\,,
\label{series}
\eea
where the redshift $z$ is now the source's redshift.
This result agrees with the Mattig result (\cite{MW}),
\be
D^M_{\ell}= r R_0 (1+z) ={c\over
H_0q_0^2}\left\{q_0z+(q_0-1)\left(\sqrt{1+2q_0z}-1\right)\right\},
\label{Mattig}
\ee
when $\rho_I= 0$.

If $b$ can be neglected the difference between the standard Mattig
result (\ref{Mattig})  and (\ref{series}) [or the exact result
(\ref{answerL})] is due to the diminished converging effect of the
mass remaining in the beam, i.e., the $(\rho_D/\rho_F){\cal R}_F$
term in (\ref{Af}). The standard result assumes all matter is
uniformly distributed in and out of the beam. If some fraction
$\rho_I/\rho_F$ exists only out of the beam where it acts only
through a negligible $b^3$ term, then a diminished fraction
$\rho_D/\rho_F$ remains to converge the beam.  In the Appendix we
conclude that for a large majority
of observations the $b^3$ term can be neglected.
We now explicitly give the analytic solution of (\ref{Af})
when $b=0$.

\section{The Analytic Solution For \hbox{{\it A}}(${\lowercase {z}}$)
when ${\lowercase {b}}=0$ }
\label{sec:b=0}

The solution of (\ref{Af}) for $b=0$ and $\rho_D=0$
was given by \cite{DC1}
and is the  limiting case ($\rho_D\rightarrow 0$) of result
(\ref{answerL}) below.
We first start by rewriting  (\ref{Af}) when $b=0$ as
\be
(1+z)^3\sqrt{1+2q_0z}\,{d\ \over dz}(1+z)^3\sqrt{1+2q_0z}\,{d\ \over
dz}\sqrt{A(z)}
+{ \rho_D\over \rho_F} 3q_0(1+z)^5\sqrt{A(z)}=0.
\label{Ab0}
\ee
This equation\footnote[3]{See the Appendix A of Seitz \& Schneider 1994 for the
first presentation of this material including a comparison of all previously
known limiting solutions.} can next be put into a recognizable self-adjoint
form
by changing  the independent
variable from $z$ to $\zeta$ and the dependent variable from
$\sqrt{A(z)}$ to  $P(\zeta)$,

\bea
1+z &\equiv& \left(1-{1\over 2q_0}\right)(1-\zeta^2)\ ,\nonumber\\
 (1+z)^2\sqrt{A(z)} &\equiv& (1-\zeta^2)P(\zeta).
\eea
For $q_0 < 1/2$ we have $\zeta^2 > 1$ and for  $q_0 > 1/2 ,\ \zeta$
is imaginary.
The resulting equation
\be
(1-\zeta^2){d^2P\over d \zeta^2} -2 \zeta\ {dP\over d\zeta} +\left[6\
{\rho_I\over\rho_F} - {4\over 1-\zeta^2}\right] P = 0\ ,
\label{aLe}
\ee
has solutions which are arbitrary linear combinations of the
associated Legendre functions of the first and second kind  (\cite{AM})
i.e.,
\be
P(\zeta) =  c_1 P^2_{\nu}(\zeta) + c_2 Q^2_{\nu}(\zeta)\ ,
\label{aLf}
\ee
where
\be
\nu \equiv \left(\sqrt{1+24\rho_I/\rho_F}-1\right)/2\, ,
\ee
 is the positive root of
 $\nu(\nu+1) = 6\rho_I/\rho_F$. The range for $\nu$ is from  $\nu= 0$
(usual pure dust FRW)
where no clumps are present, to $\nu=2$ where all mass is in the form
of clumps
and no dust is present.

Boundary conditions can now be chosen  appropriate for either
luminosity distance $D_{\ell}$ as above or for apparent-size/angular
distance $D_{\angle}$\,. For apparent-size distances $\sqrt{A}=\delta
\theta\,D_{\angle}$\,, the boundary values are specified at the
observer, $ D_{\angle}|_0=0$
and $d D_{\angle}/d v|_0=-1$ (or equivalently $d D_{\angle}/d
z|_0=c/H_0$). For the $b=0$ case, the two types of
solutions are easily related by the Wronskian of (\ref{Ab0}) (see
\cite{EI}),
\be D_{\ell}(z)=D_{\angle}(z)(1+z)^2.
\label{Etherington}
\ee
Now switch to the more easily implemented  apparent-size boundary
conditions and hence require:
\bea
\left( c_1 P^2_{\nu}(\zeta) +  c_2
Q^2_{\nu}(\zeta)\right)\Big\vert_{z = 0} &=& 0 \ ,\nonumber\\
{q_0\over\sqrt{1-2q_0}}\left( c_1{d P^2_{\nu}(\zeta)\over d\zeta} +
c_2  {d Q^2_{\nu}(\zeta)\over d\zeta}\right)\Big\vert_{z = 0} &=&
{c\over H_0}\ .
\label{boundary}
\eea
These two equations can be used to find $c_1$ and $c_2$. The results
then
simplify by using the constancy of the Wronskian of (\ref{aLe}),

\be
W[P^2_{\nu}(\zeta),Q^2_{\nu}(\zeta)] \equiv P^2_{\nu}(\zeta){d
Q^2_{\nu}(\zeta)\over d \zeta} -Q^2_{\nu}(\zeta){d
P^2_{\nu}(\zeta)\over d \zeta} =
{(\nu+2)(\nu+1)(\nu)(\nu-1)\over (1-\zeta^2)}\ ,
\label{Wronskian}
\ee
to give
\bea
c_1 &=& {c\over H_0}
{
1\over \sqrt{1-2q_0}
}
{
2\over (\nu+2)(\nu+1)(\nu)(\nu-1)
}
Q^2_{\nu}\left({1\over\sqrt{1-2q_0}}\right)
\ ,\nonumber\\
c_2 &=& -{c\over H_0}
{1\over \sqrt{1-2q_0}}
{
2\over (\nu+2)(\nu+1)(\nu)(\nu-1)
}
P^2_{\nu}\left({1\over\sqrt{1-2q_0}}\right)\ .
\eea
The resulting expression for apparent-size distance is
\bea
&&D_{\angle}(q_0, \nu ; z)= {c\over H_0}{1\over\sqrt{1-2q_0}}
{
2\over (\nu+2)(\nu+1)(\nu)(\nu-1)
}
\times
\label{answerL}
\\
&&\left({1\over
1+z}\right)\left[Q^2_{\nu}\left({1\over\sqrt{1-2q_0}}\right)
P^2_{\nu}\left({\sqrt{1+2q_0z}\over\sqrt{1-2q_0}}\right)-
P^2_{\nu}\left({1\over\sqrt{1-2q_0}}\right)
Q^2_{\nu}\left({\sqrt{1+2q_0z}\over\sqrt{1-2q_0}}\right)\right].\nonumber
\eea
This result is valid for all $q_0 \ge 0$ but some care has to be
taken with
the limit $q_0\rightarrow 1/2$ as well as with the $\nu\rightarrow
0,1$ limits.
The $\nu\rightarrow 0$
limit ($\rho_I=0$) is the Mattig result for FRW and is related to
(\ref{Mattig})
by (\ref{Etherington}). The $\nu = 2$\  (i.e., the  $\rho_D=0$ Dyer \& Reoder
1972 result)
is also obtained from (\ref{answerL}) [see \cite{SS}].

Because the associated Legendre equation (\ref{aLe}) is a special
case of the
hypergeometric equation the above solution (\ref{answerL}) can be
written in terms of hypergeometric functions.\footnote[1]{That
differential equation (\ref{Af}) was of the  hypergeometric type when
$b=0$ was recognized by \cite{DC2}.}
 This makes the analytic continuation to $q_0\ge 1/2$ transparent and
gives a resulting expression for $D_{\angle}$ (or $D_{\ell}$) whose
constituent parts no longer have singularities. Computer routines can
then be used to evaluate these distances without having to avoid the
singularities at $q_0=1/2,\, \nu= 0,$ and $\nu=1$.
The two identities needed are:\footnote[2]{Note that in  8.1.5 of
\cite{AM} a $\pi^{-1/2}$ is missing.}

\bea
P^2_{\nu}(\zeta)&=&{\Gamma(-{1\over2}-\nu)\zeta^{-\nu+1}\over
2^{\nu+1}\pi^{1/2} (\zeta^2-1)}
 \ {}_2F_1\left({\nu\over2}-{1\over2}, {\nu\over2}; \nu+{3\over2};
{1\over \zeta^2}\right)\nonumber\\
&+&{2^{\nu}\Gamma({1\over2}+\nu)\zeta^{\nu+2}\over \pi^{1/2}
\Gamma(\nu -1)(\zeta^2-1)}
\ {}_2F_1\left(-{\nu\over2}-1, -{\nu\over2}-{1\over2}; {1\over2}-\nu;
{1\over \zeta^2}\right),
\eea
\bea
Q^2_{\nu}(\zeta)&=&{\pi^{1/2} \Gamma(\nu+3)\zeta^{-\nu+1}\over
2^{\nu+1}\Gamma(\nu+{3\over2})(\zeta^2-1)}
\ {}_2F_1\left({\nu\over2}-{1\over2}, {\nu\over2}; \nu+{3\over2};
{1\over \zeta^2}\right)\nonumber\\
&=&{\pi^{1/2} \Gamma(\nu+3)\zeta^{-\nu-3}(\zeta^2-1)\over
2^{\nu+1}\Gamma(\nu+{3\over2})}
\ {}_2F_1\left({\nu\over2}+2, {\nu\over2}+{3\over 2}; \nu+{3\over2} ;
{1\over \zeta^2}\right),
\eea

giving the second form of the solution to equation (\ref{Ab0}),
\bea
&&D_{\angle}(q_0, \nu ; z)={c\over H_0}{1\over (\nu+{1\over2})}\times
\label{answerH}
\\
&&
\Biggl[
{(1+2q_0z)^{1+\nu/2}\over (1+z)^2}
\ {}_2F_1\left({\nu\over2}+2, {\nu\over2}+{3\over 2}; \nu+{3\over2} ;
1-2q_0\right)
{}_2F_1\left(-{\nu\over2}-1, -{\nu\over2}-{1\over2}; {1\over2}-\nu;
{1-2q_0\over 1+2q_0z}\right)
\nonumber\\
&-&{1\over
(1+2q_0z)^{3/2+\nu/2}
}
\ {}_2F_1\left(-{\nu\over2}-1, -{\nu\over2}-{1\over2}; {1\over2}-\nu;
1-2q_0\right)
{}_2F_1\left({\nu\over2}+2, {\nu\over2}+{3\over 2}; \nu+{3\over2} ;
{1-2q_0\over 1+2q_0z}\right)
\Biggr].\nonumber
\eea
This form is well behaved as a function of $\nu$ as well as  in the
neighborhood of $q_0=1/2$, and
is needed beyond the circle  of convergence ($|\xi|  = 1$) of the
hypergeometric series ${}_2F_1(a,b;c;\xi)$ only on
the negative real axis $\xi = x \le -1$ when $q_0 \ge 1$ (where it is
easily continued).
In Fig. 2,\,  $D_{\angle}$ is plotted as a function of $z$ for three
values of $\nu$ and
three values of $q_0$ and in Fig. 3 the related magnitude-redshift curves are
plotted. The reader should recall that apparent (bolometric) magnitudes $m$
are related to absolute (bolometric) magnitudes $M$ and luminosity distances
$D_{\ell}$ [see (\ref{Etherington})] by
\be
m=M+5\log_{10}(D_{\ell}/10 pc)=5\log_{10}({H_0\over c} D_{\ell})+constant,
\label{mz}
\ee
which when evaluated as a function of redshift is called the $m$--$z$
(magnitude-redshift) relation. As pointed out in the next section the $constant
\equiv M-5\log_{10}({H_0\over c} 10 pc)$ turns out to be observationally
independent of the Hubble parameter.
In the next section we investigate the sensitivity of this relation to the
clumpiness parameter $\nu$.
\section{Application}
\label{sec:conclusions}

We conclude by illustrating the effect of inhomogeneities on the
determination of $q_0$ when Type Ia supernovae (SNe Ia) are used in the Hubble
diagram. Such a determination  is expected to be feasible in the near future.
Attempts to use
galaxies or quasars in this way are foiled by the intrinsic dispersion in
their luminosities and by uncertainties associated with their evolution.
However, normal SNe Ia (those that have nonpeculiar spectra, light--curve
shapes, and broad band colors) are luminous point sources that form a well
defined ridge line in a Hubble diagram (for recent reviews of SNe Ia see
\cite {BD}  and references therein).  Even just the application of
a simple B--V color criterion defines a subsample of SNe Ia that has mean
absolute magnitudes $\rm M_B \simeq M_V \simeq -19.8 \pm 0.1 + 5 log (H_0/50)$
and dispersions about the means $\rm \sigma(M_B) \simeq \sigma(M_V) \simeq 0.2$
magnitudes (\cite{VT1,HM,VT2}). With sufficient spectroscopic or photometric
information it may
be possible to control the individual SN Ia relative magnitudes to 0.1 or even
better (\cite{FA,HM,RA,TG}).  The value of $H_0$, and thus the calibration of
the
mean SN Ia absolute magnitude, does \underbar{not} enter the determination of
$q_0$ from
the Hubble diagram.

In the standard SN Ia model, a mass accreting white dwarf approaches the
Chandraskhar limit, ignites degenerate carbon near its center, undergoes a
thermonuclear instability, and disrupts completely.  The light curve is
powered by the radioactive decay of freshly synthesized $^{56}$Ni through
$^{56}$Co to $^{56}$Fe.  Thus, to first order, the peak absolute magnitude of
SNe Ia is not expected to vary with cosmic time.  Observationally, some
evidence for a dependence of the properties of SNe Ia on the age of the
stellar population at their sites has been found (\cite{BD2,HM}).  This
constitutes evidence for mild evolution of the
SN Ia population with cosmic time, but it implies that SNe Ia at high
redshift, when the universe was younger, ought to have an even smaller
dispersion among their properties than do the nearer SNe Ia that constitute
the bulk of the present observational sample.  In any case, with sufficient
spectroscopic and photometric information it should be possible to identify
counterparts of the high redshift SNe Ia among the nearer SNe Ia.

Recently it has been demonstrated that SNe Ia at cosmologically significant
redshifts can be discovered and measured. \cite{PS2} has
announced the discovery of SN 1994F, a SN Ia at $z = 0.354$; SN 1994G, a
probable SN Ia at $z = 0.425$; and SN 1994H, a probable SN Ia at $z = 0.373$.
\cite{PS1}  present a light curve of SN 1992bi, a probable SN Ia
at z=0.458, which they use with the standard Mattig relation to make a
preliminary estimate of the value of $q_0$. Recently additional remote SNe Ia
and probable SNe Ia have been discovered (C. Pennypacker, personal
communication).

Fig. 4, which is an enlarged portion of Fig. 3 near the redshift $z = 0.458$
of SN 1992bi, shows that clumping can have a significant effect on the
determination of
$q_0$ at this redshift.  In Fig. 5, the quantity $\rm 5\, log_{10}$
$(H_0/c)D_l(q_0,\nu; z=0.458)$ [see (\ref{mz})] is plotted as a function of
$q_0$ for three
values of the clumping parameter $\nu$.  For example, if a given set of
observations gives $q_0 = 0.1$ (\cite{PS1}) when the Mattig relation is used,
the $\nu
= 2$ (completely clumpy, $\rho_I = \rho_F$) relation would give $q_0 = 0.15$;
that is, if the universe is completely clumpy then neglecting to take
clumpiness
into account causes an underestimate of $q_0$ by about 50 percent. The
difference between $q_0 = 0.1$ and $q_0 = 0.15$ is significant quantitatively,
but not qualitatively.  On the other hand, if the observations were such that
the Mattig relation gave $q_0 = 0.34$, then the $\nu = 2$ relation would give
$q_0 = 0.51$.  This difference is more than merely quantitative --- it is the
difference between an open and a closed universe.

We find that it is just when redshifts are large enough for the
determination of $q_0$ that the effects of clumps become significant.  Figs.
4 and 5 show that if $q_0$ were known, distinguishing between a completely
clumpy universe and a universe with no clumps would require that the average
apparent
magnitude of a set of remote SNe Ia be compared to that of nearby SNe Ia to an
accuracy $\delta m \simeq 0.05$ magnitude. Such statistical accuracy should
soon
be available
at the rate SNe Ia are being discovered.  In
practice, though, the $m$-$z$ relation only constrains the $q_0$ and $\nu$
parameters in combination.  Independent information must be used to determine
their individual values.

\acknowledgements
This work was supported
by the Department of Energy, the Southern Association for High
Energy
Physics (SAHEP) funded by the Texas National Research Laboratory
Commission (TNRLC), and by NSF grant AST 91-15061.

\appendix
\section{Can the parameter ${\lowercase {b}}$ be
neglected?}
\label{sec:neglect}
Based on the series solution (\ref{series}), \cite{KR} concluded that
the $b^3$ term appearing in (\ref{Af}) and (\ref{A3}) would be
insignificant for  observations of large objects such as galaxies.
Further arguments are now given to reassert that the probability
of the $b^3$ term affecting an observation is rather small. A priori,
observing a significant effect of $b$ on the magnitude-redshift
relation  should be about as likely as a single observation resulting
in a gravitational lens effect.
When either lensing or a significant distortion of a light beam occurs,
the cause is the same conformal curvature term in the gravity field.
Rauch (1991) has come to a similar
conclusion about the unlikely probability of observing gravitationally
lensed SNe Ia at small z as we do here.
In order to estimate  the significance of focusing by external mass clumps on
a passing beam of light, i.e., to estimate the relative significance
of $b^3$ in (\ref{A3}), a comparison of the magnitudes of the two terms
that appear as coefficients of $A$ can be made.
Even if such a comparison implies that the $b^3$ term is significant
compared to the $\cal R'$ term, they both might be unimportant
compared to the $z$ development caused by the $A'''$ or $A'$ terms.
The ratio of the two $A$ terms is
\be
{\rho_I\over\rho_D} {b^3(1+z)^6\over\cal R'}=
{\rho_I\over\rho_D} \left({b^3\over 15 q_0(H_0/c)^3}\right)
{1\over (1+z)\sqrt{1+2q_0z}}\, .
\ee
The unitless combination

\be
B_0\equiv {b^3\over 15 q_0(H_0/c)^3}=
{4\over 5} {c\over H_0}
{\langle(G m/c^2)^{4/3}\rangle \over \langle(G m/c^2)\rangle }
\left[{(G m/c^2)^{1/3}\over a}\right]^2,
\label{B15}
\ee
is clearly the parameter determined by the structure of the
inhomogeneities, which affects the relative importance.
In the series solution (\ref{series}), $B_0$ first appears in the
$z^4$ term, already indicating that it is probably not important at
small redshifts.

\cite{DC3} numerically integrated (\ref{A3}) for
numerous values of ${\rho_I/\rho_D} $ and $B \equiv 15 B_0$. From
their graphs we can  conclude that as long as  $B_0 \sim 1$  and
observations are at redshifts $z < 1$ then the $b^3$ term
is not significant. In Fig. 6 the recomputed numerical solutions of
(\ref{A3}) are given for values of $B_0= 0, 1$, and $5$, $q_0 = 0.1$
and $0.5$, and $\rho_I/\rho_F = 0$ and $1$, confirming the above
conclusion.
For example for $ q_0 = 0.5$ the magnitude of a SN Ia at $z \approx
0.5$ in the standard dust filled FRW universe would appear 0.075
magnitudes brighter than in the $B_0 = 0$ clumpy universe. If, in the
clumpy universe, $B_0 = 1$ or $B_0 = 5$ the magnitude would be
brighter by 0.01 and 0.05 magnitudes respectively. For $q_0 = 0.1$
all changes are about 1/4 of those for $q_0=0.5$\,.
The conclusion is that for $B_0 \sim 1$ (and smaller) the $b^3$ term
makes only a small increase in intensity ($\sim$ 1/7 or less)
compared to the decrease in intensity due to the missing matter in
the light beam ($\nu = 2$ rather than $\nu = 0$).
As to what $B_0$ values are known to exist:  for globular clusters,
$B_0\sim 10^{-2}$; for galaxies and galaxy clusters, $B_0\sim 1$; for
stars, $B_0\sim 10^9$; and for black holes at the centers of spiral
galaxies, $B_0\sim 10^{25}$.

What is now shown is that the probability of seeing the effects of a
$B_0$ larger than $\sim 1$ is quite small (for these limited
redshifts).
If it is assumed that there is a randomly distributed population of
essentially point inhomogeneities  of number density $n=n_0R_0^3/R^3$
between us and an observed object at redshift $z$, the probability
that the observed light beam comes
no closer than a distance $a'$ to the center of any of the
intervening masses
can be computed by evaluating
\be
P(r) = \exp\left\{-\int_0^r{n \pi a'^2 R dr\over \sqrt{1-k
r^2}}\right\}\, .
\label{probr}
\ee
Assuming the clumps are points will maximize their possible effects.
The parameter $r$ used here is the FRW comoving radial coordinate
centered on the source and $R$ is the FRW scale factor.
This result follows from the simple assumption that $-\Delta P\equiv
P(r)-P(r+\Delta r)$, the probability a beam make it to $r$ having had
no impact parameter less than $a'$ and  then having an impact
parameter less than $a'$ within the next $\Delta r$, is  given by
\be
-\Delta P = P(r)\ { \pi a'^2\times  n\ 4 \pi r^2 \Delta r R^3\over
4\pi r^2 R^2 \sqrt{1-k r^2}}\, .
\ee
This is a simple ratio of the sum of impact disk areas in a spherical
shell of thickness $R\Delta r/\sqrt{1-k r^2}$ to the total area on
the sky, at comoving distance $r$.
After changing from co-moving coordinate $r$ to redshift $z$,
(\ref{probr})
can be easily integrated to give
\be
P(z) = \exp\left\{-{\pi a'^2 n_0 c\over  H_0} f(z,q_0)\right\}\, ,
\label{probz1}
\ee
where
\be
 f(z,q_0)= {1\over 2 q_0^2}\left\{
{\left[(1+2q_0z)^{3/2}-1\right]\over 3}
-(1-2q_0)\left[(1+2q_0z)^{1/2}-1\right]\right\}\, .
\ee
If it is assumed that all members of this population of
inhomogeneities have the same
mass $m$, (\ref{probz1}) can be rewritten as a function of the
unitless parameter $B'_0$ of (\ref{B15}) as
\be
P(z) = \exp\left\{-{3 q_0\over 5 B'_0}{\rho_I\over \rho_F}
f(z,q_0)\right\}\, .
\ee
$P(z)$ is interpreted as the probability that all impact parameters
($a$ values as the light beam interacts with inhomogeneities on its
path from a source at $z$ to us) have values such that $4 G m /(5 H_0
c a^2) = B_0$ is less than  $B'_0$.
In Fig. 7, $P(z)$ is plotted for several values of $q_0$ and
$(\rho_I/\rho_F) B_0'$. For a given $B_0'$ the maximum probability occurs
when $\rho_I/\rho_F=1$, i.e., when all matter is in point clumps.
{}From Fig. 7 it is clear that for redshifts $z < 1$,  the
probability of
having a value of $B_0 < 1 $ is high. For such values of $B_0$ it has
already
been concluded that the $b$ term in (\ref{A3}) has a rather small
effect. What is expected to happen is that an occasional observation
of a SN Ia will occur
where the intensity is increased because it happens to be closely
aligned with  a dark and unseen  interveining galaxy. According to
Fig. 7 this can't happen frequently. However, when it does, the SN will
either sit far from the $m$-$z$ curve where it can be excluded, or sit
close to the
$m$-$z$ curve and simply
increase the  error in the estimated value of $q_0$.

\eject

\begin{figure}
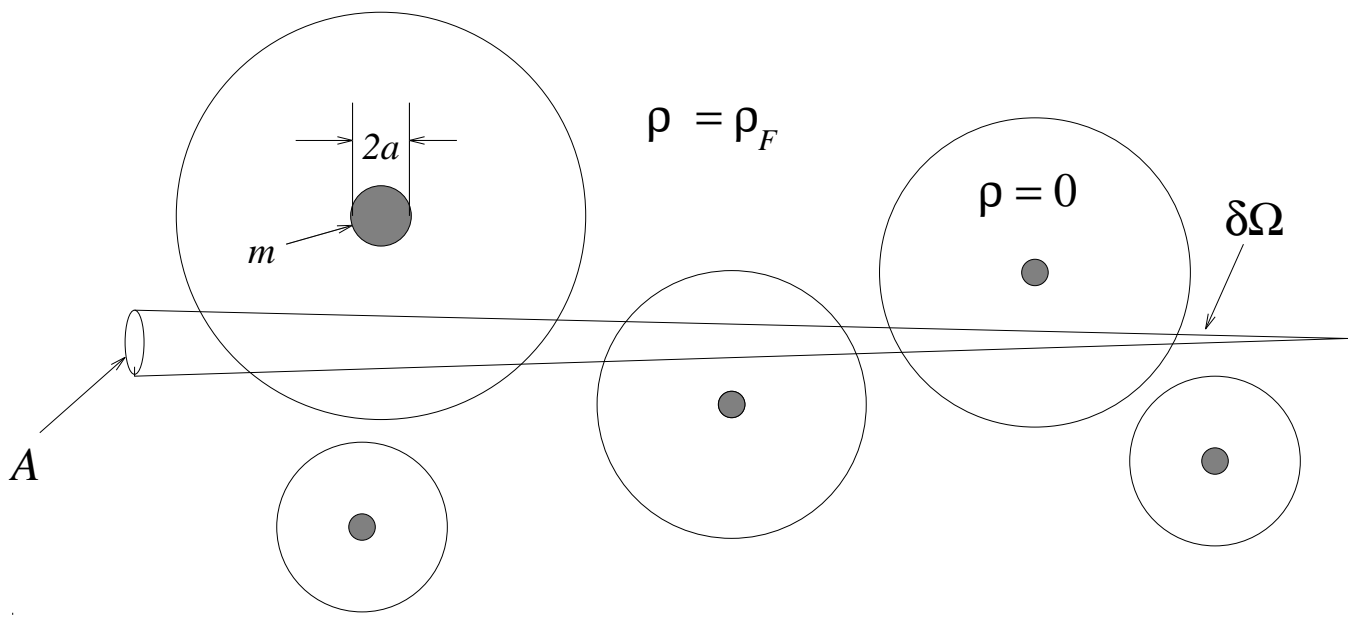

\caption{Radiation beam of cross section $A$ propagating
through a Swiss-cheese universe from distant source to observer.}
\label{Fig1}
\end{figure}

\begin{figure}
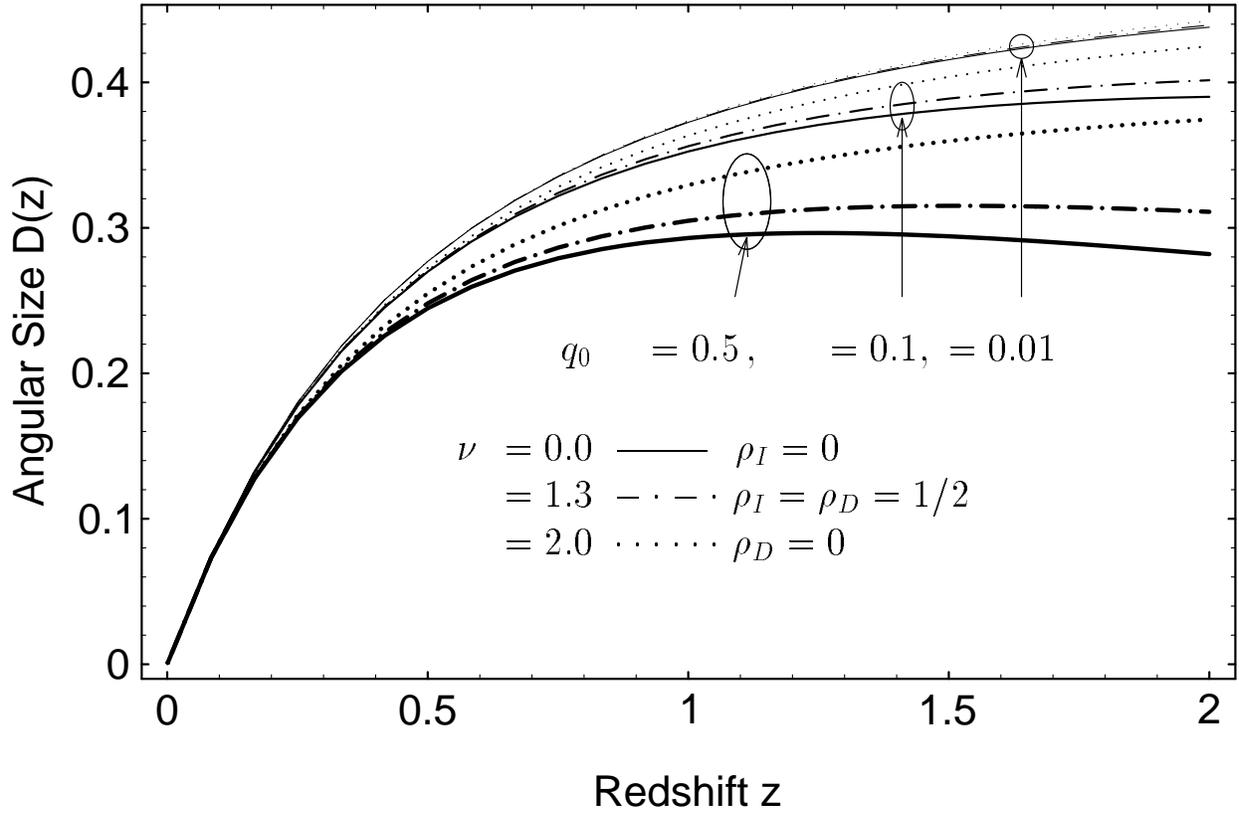

\caption{Apparent-size distance ${H_0\over c}D_{<}(q_0,\nu;z)$
as a function of redshift $z$ for three values of $q_0$ and
three values of $\nu$. }
\label{Fig2}
\end{figure}

\begin{figure}
\caption{Magnitude-redshift relation,
$5\log_{10}{H_0\over c}D_{\ell}(q_0,\nu;z)$,
as a function of redshift $z$ for three values of $q_0$ and
three values of
$\nu$. }
\label{Fig3}
\end{figure}

\begin{figure}
\caption{Magnitude-redshift relation,
$5\log_{10}{H_0\over c}D_{\ell}(q_0,\nu;z)$ ,
as a function of redshift $z$ for three values of $q_0$ and
three values of $\nu$. }
\label{Fig4}
\end{figure}

\begin{figure}
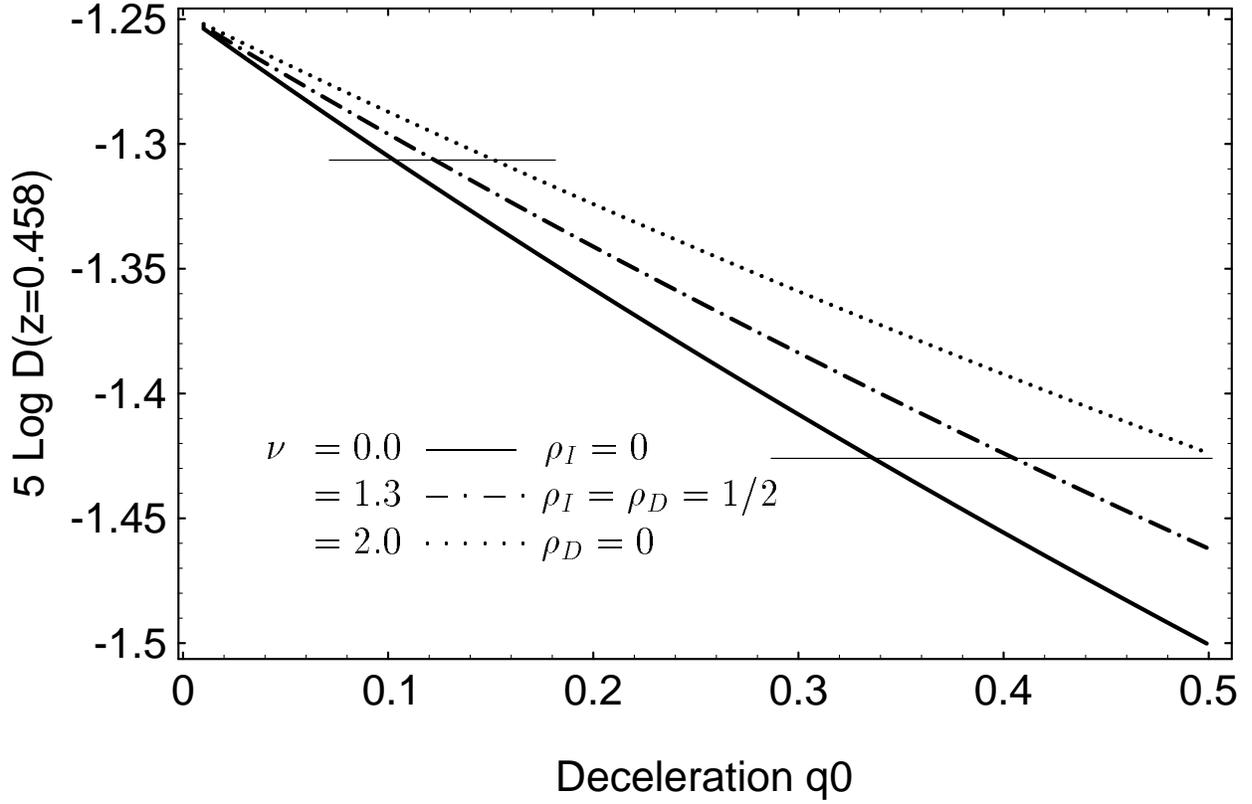

\caption{Magnitude-redshift relation,
$5\log_{10}{H_0\over c}D_{\ell}(q_0,\nu;z=0.458)$,
as a function of the deceleration parameter $q_0$ for three values
of $\nu$.}
\label{Fig5}
\end{figure}

\begin{figure}
\caption{Magnitude-redshift relation,\,$5\log_{10}{H_0\over c}D_{\ell}
(q_0,\nu,B_0;z)$,\,as a function of redshift $z$ near 0.458 for
two values of $q_0$ and $\nu$,\,and three values of $B_0$.}
\label{Fig6}
\end{figure}

\begin{figure}
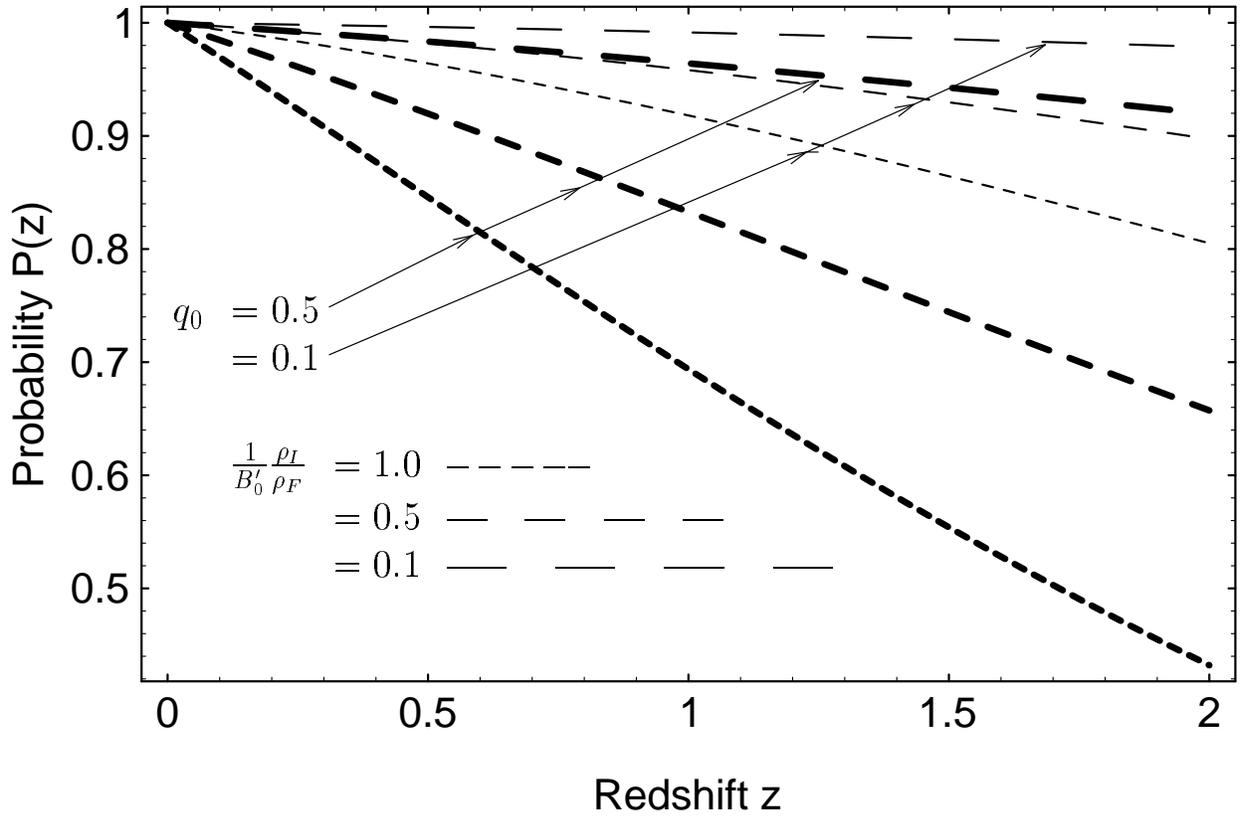

\caption{Probability of $B_0 < B_0'$ in a universe of given $\rho_I/\rho_D$}
\label{Fig7}
\end{figure}

\end{document}